\documentclass[a4paper,11pt]{article}
\usepackage[utf8]{inputenc}
\usepackage[english]{babel}
\usepackage{braket}
\usepackage{xspace}
\usepackage{bbm}
\usepackage{mathdots}
\usepackage{stackrel}
\usepackage[dvipsnames]{xcolor}
\usepackage{appendix}
\usepackage{hyperref}
\hypersetup{
 colorlinks=true,
 linkcolor=blue,
 anchorcolor = blue,
 citecolor = blue,
 filecolor = blue,
 urlcolor = blue
}

\usepackage{tikz}
\usetikzlibrary{decorations.markings}
\usepackage{pgfplots}

\usepackage{mathtools}
\usepackage{bbm}
\usepackage{comment}
\usepackage{subfigure}
\usepackage[T1]{fontenc}               % codifica dei font
\usepackage[left=3cm,
            right=2.5cm,
            top=2.5cm,
            bottom=3cm
            ]{geometry}                % spaziatura bordi
\usepackage{graphicx}
\usepackage{mathrsfs}
\usepackage{appendix}
\usepackage{fancyhdr}
\usepackage{amsmath,                   % simboli matematici
            amssymb,                   % altri simboli matematici
            amsthm}                    % stili teorema

\usepackage{setspace}
\usepackage{cite}
\usepackage{cancel}
\usepackage[margin=30pt, bf, font=small, center, justification=justified]{caption}[2004/07/16]

\usepackage{physics}

\newcommand{\filib}[1]{{\color{Green} FA: #1}}

\newcommand{\be}{\begin{equation}}
\newcommand{\ee}{\end{equation}}
\newcommand{\bea}{\begin{eqnarray}}
\newcommand{\eea}{\end{eqnarray}}
\newcommand{\bse}{\begin{subequations}}
\newcommand{\ese}{\end{subequations}}

\title{Multiple crossing during dynamical symmetry restoration and implications for the quantum Mpemba effect}

\author{ Konstantinos Chalas$^{1}$, Filiberto Ares $^{1}$,  Colin Rylands$^{1}$ \&  Pasquale Calabrese$^{1,2}$ }

\date{}

\begin{document}

\maketitle
{\small
\vspace{-5mm}  \ \\
{$^{1}$}  SISSA and INFN Sezione di Trieste, via Bonomea 265, 34136 Trieste, Italy\\[-0.1cm]
\medskip
{$^{2}$}  International Centre for Theoretical Physics (ICTP), Strada Costiera 11, 34151 Trieste, Italy\\[-0.1cm]
\medskip
}

\begin{abstract}
Local relaxation after a quench in 1-D quantum many-body systems is a well known and very active problem with rich phenomenology.
Except for pathological cases, the local relaxation is accompanied by the local restoration of the symmetries broken by the initial state that are preserved by the unitary evolution.  Recently, the entanglement asymmetry has been introduced as a probe to study the interplay between symmetry breaking and relaxation in an extended quantum system. In particular, using the asymmetry, it has been shown that the more a symmetry is initially broken, the faster it may be restored. This surprising effect, which has been also observed in trapped-ion experiments, can be seen as a quantum version of the Mpemba effect and is manifested by the crossing at a finite time of the entanglement asymmetry curves of two different initial symmetry breaking configurations.
In this paper we show,  how, by tuning the initial state, the symmetry dynamics in free fermionic systems can display much richer behaviour than seen previously. In particular, for certain classes of initial states, including ground states of free fermionic models with long-range couplings, the entanglement asymmetry can exhibit multiple crossings. This illustrates that the existence of the quantum Mpemba effect can only be inferred by examining the late time behaviour of the entanglement asymmetry. 
\end{abstract}

\section{Introduction}
\label{sec:introduction}

Understanding if and how,  closed, many-body,  quantum systems relax is a question of both fundamental and practical importance to modern theoretical physics.  This topic has been investigated for almost as long as the existence of quantum theory, but has received further impetus in recent decades through the development of many state of the art experimental and theoretical techniques~\cite{pssv-11, ge-16, cem-16}.  During this time there have been numerous seminal contributions to the field, providing insight into the mechanisms of relaxation and thermalization of closed systems,  as well as the obstructions to these processes.  A straightforward but natural question to ask  in this context, which has been curiously overlooked until recently, is: Given two non-equilibrium states, which one will relax faster, the one closer to equilibrium or the one further from it.  Surprisingly, the answer to this question can be either yes or no, depending on the particular setup.  The former scenario is in line with our intuitive understanding of systems which are close to equilibrium, however the latter subverts this and is reminiscent of the famous Mpemba effect of classical physics~\cite{cool, clathrate, polymers, carbon, granular, cold, raz, raz2, kumar, bechhoefer, kumar2, wv-22, teza2023relaxation, wbv-23, bwv-23}.  This anomalously fast relaxation of a far from equilibrium quantum state was recently predicted to occur for tilted ferromagnetic states evolved unitarily using the XX spin chain~\cite{ASYMM00asymmetry_og}.  Owing to its similarities to the classical case, but being a distinctly quantum phenomenon, it was  dubbed the quantum Mpemba effect (QME).  The phenomenon was later studied in several other scenarios, including both free~\cite{ASYMM01murciano2023xympemba}, interacting integrable~\cite{ASYMM08rylands2023microscopic, ASYMM02bertini2023dynamics} and chaotic models~\cite{liu_mpemba_circuit-24,tcd-24} as well as in free dissipative systems~\cite{cma-24} and higher dimensions~\cite{yac-24}.  For systems which admit a quasi-particle description, the
underlying mechanisms for the QME have been determined and a set of predictive criteria related to transport properties have been proposed~\cite{ASYMM08rylands2023microscopic}. Subsequently, it has been observed experimentally~\cite{ASYMM07joshi2024observing} in a system of trapped ions using the randomized measurement toolbox~\cite{elben}. Parallel to this, there have been several recent studies, both theoretical~\cite{quantum1, quantum2, quantum3, quantum4, quantum5, quantum6, cth-23-2, spc-24, mczg-24} and experimental~\cite{shapira-24, zhang-exp-24}, of the classical Mpemba effect in quantum systems.  Although similar in spirit to the QME,  the latter phenomenon occurs in few body systems at finite temperature or coupled to an external reservoir.  Following the nomenclature of phase transitions, we thus describe these cases as the Mpemba effect in a quantum system as opposed to a quantum Mpemba effect, which is driven purely by quantum rather than thermal fluctuations. 

Although it can be witnessed through numerous different observables, an elegant quantity which captures the QME for systems with a conserved charge evolved from symmetry broken states is the entanglement asymmetry (EA)~\cite{ASYMM00asymmetry_og}.    This is an entanglement based measure which serves as an ideal tool for investigating intricate symmetry breaking and restoration at the level of a subsystem, in equilibrium and non-equilibrium extended systems alike.  Simply stated,  the EA quantifies the distance a state is from being symmetric; it is finite and positive in symmetry broken states and zero when a symmetry is present.  Thus,  for non-equilibrium states which break a symmetry but which, under time evolution restore it,  the EA (typically) evolves as a monotonically decreasing function of time.   In this setting, the QME is witnessed by a single crossing of these curves for different initial states.  
This fact might suggest that the observation of the quantum Mpemba effect could be  inferred from the analysis of short time dynamics only. 
To disprove this naive expectation, in this work, we investigate the dynamics of the entanglement asymmetry from a class of initial states in free-fermion chains which display richer relaxation dynamics in which the entanglement asymmetry curves of different initial states exhibit multiple crossings in time.  
Consequently, the occurrence of the Mpemba effect can be understood only by a careful analysis of the long time dynamics.

\textbf{Quench dynamics and symmetry restoration:} The most common scenario for studying the non-equilibrium dynamics of closed quantum systems is the quantum quench.  In this setup, we consider an initial state $\ket{\Psi_0}$, which is then unitarily evolved according to a Hamiltonian $H$, having the time evolved state
\be
\label{eq:time_ev_state}
    \ket{\Psi(t)}=e^{-i H t}\ket{\Psi_0}.
\ee
The state $\ket{\Psi_0}$ is not
an eigenstate of $H$ and generically has non-zero overlap with exponentially many eigenstates of $H$. Since the full system evolves unitarily, no relaxation can happen at the level of the full system. 
Instead, we study a finite subsystem, $A$,  of  size $|A|=\ell$, through its reduced density matrix $\rho_{A}(t)$, after tracing out its complement $\bar{A}$,
\be
\label{eq:RDM_finite_time}
    \rho_{A}(t)=\operatorname{Tr}_{\bar A}\rho(t)=\operatorname{Tr}_{\bar A}\ketbra{\Psi(t)}{\Psi(t)}.
\ee
On timescales  much longer than those associated to the subsystem size, the subsystem is expected to relax to a stationary state.    If we quench with free fermionic Hamiltonians, which are integrable, the stationary state is not thermal, i.e.  a Gibbs ensemble, but is instead a generalized Gibbs ensemble (GGE), where aside from the Hamiltonian, we have to include the constraints imposed by all the constants of motion  and set by the initial state~\cite{rdyo-07, cef-12, vr-16, ef-16}.

We are particularly interested in quench Hamiltonians $H$ possessing a $U(1)$ symmetry that corresponds to a local conserved charge $Q$ which can be written as a sum of terms in $A$ and $\bar A$, i.e.  $Q=Q_{A}+Q_{\bar{A}}$.
The state of the system $\ket{\Psi(t)}$ is symmetric if  $\left[\rho(t),Q\right]=0$.  If we trace out the complement of the subsystem $A$, then $\left[\rho_{A}(t),Q_{A}\right]=0$, meaning that, for a symmetric state, $\rho_A(t)$ is block-diagonal in the eigenbasis of $Q_A$, so it has no off-diagonal elements connecting different charge subsectors of $Q_{A}$. For such symmetric states, a question
that has been intensively studied in recent years is how entanglement
is distributed into the charge subsectors of $Q_{A}$~ \cite{SREE01Parez_2021,SREE02Piroli_2022,SREE03_fcs_sree_st_duality,SREE04Murciano_dissipation_2023,SREE05_mbl_entan,SREE06SierraEquipartition,SREE07GoldsteinSela, SREE12Bonsignori_2019,SREE13murciano_di_Giulio_2020,SREE14_qpp_sree_cft_ffs,SREE15Laflorencie_2014, SREE16_noisy_q_comps, SREE17_vitale_sree_det, SREE18_vitale_purification,SREE19_ent_barrier}. As a result of these, it has been shown that, at leading order, entanglement entropy is usually equipartitioned between charge sectors. Here instead, we are interested in initial states which do not preserve the symmetry generated from this conserved charge,  $\left[\rho(0),Q\right]\neq0$, while also satisfy $\left[\rho_{A}(0),Q_{A}\right]\neq0$.

We consider the dynamics emerging from such symmetry breaking states with respect to a Hamiltonian that respects this symmetry.  
Generally, in such circumstances the  dynamics will locally restore the symmetry at the level of the subsystem~\cite{ASYMM00asymmetry_og}, meaning that $\lim_{t\rightarrow\infty}\left[\rho_{A}(t),Q_{A}\right]=0$. For one-dimensional systems, this can be understood through the Mermin-Wagner theorem. A generic system will relax locally to a finite temperature Gibbs ensemble, or to a generalized Gibbs ensemble if it is integrable, which is precluded from breaking any continuous symmetry. 
Exceptional cases exist, in both free and interacting systems, where this does not happen, e.g. if the initial state ignites non-Abelian charges present in the GGE steady state \cite{ASYMM06lackofrestoration}, however that shall not be the case for us.  
With this in mind,  we see that symmetry restoration can be used as a coarse-grained proxy of relaxation to the steady state as stated in Ref.~\cite{ASYMM08rylands2023microscopic}.

As mentioned above,   symmetry breaking can be quantified via the recently introduced entanglement asymmetry  \cite{ASYMM00asymmetry_og}, which measures how far away a state is from being symmetric.  This is defined as
\begin{equation}
\label{eq
:EA_def}
    \Delta S_{A}(t)=S(\rho_{A,Q}(t))-S(\rho_{A}(t)),
\end{equation}
where $S(\rho)=-\operatorname{Tr}\left[\rho \log \rho\right]$ is the von Neumann entropy of $\rho$ and 
\begin{equation} 
\label{eq:symmetrized_RDM_raq}
\rho_{A,Q}(t)=\sum_{q}\Pi_{q}\rho_{A}(t)\Pi_{q}.
\end{equation}
Here we have introduced  $\Pi_{q}$,  the projector onto the charge subsector $q$ of $Q_{A}$. 
This new reduced density matrix is thus  a symmetrized version of the state, obeying $[Q_A,\rho_{A,Q}]=0$, that alternatively can be viewed as the state of the system after a nonselective projective measurement of $Q_A$.   By using the properties of the projector, one can see that $\Delta S_{A}(t)$ is the relative entropy between our symmetry breaking state $\rho_{A}(t)$ and the symmetrized version $\rho_{A,Q}(t)$ \cite{ASYMM00asymmetry_og}. 
Being a relative entropy endows the EA with some important mathematical properties. In particular,  it is non negative, $\Delta S_A(t)\geq 0$ and, moreover, $\Delta S_A(t)=0$ if and only if $\rho_A(t)=\rho_{A,Q}(t)$.  The vanishing of $\Delta S_A(t)$ thus implies the restoration of symmetry in the subsystem and, by proxy, the relaxation to the stationary state. 

To calculate $\Delta S_{A}(t)$ we  use the replica trick, typically employed for computing entanglement measures~\cite{hlw-94, cc-04}.  To this end,  we define the order-$n$ R\'enyi entanglement asymmetries $\Delta S^{(n)}_{A}(t)$ as
\begin{equation}
\label{eq:nth_renyi_asymmetry_def}
    \Delta S^{(n)}_{A}(t)=S^{(n)}(\rho_{A,Q}(t))-S^{(n)}(\rho_{A}(t)),
\end{equation}
where $S^{(n)}(\rho)=\frac{1}{1-n}\log \operatorname{Tr}\left[\rho^{n}\right]$.  From these, we recover the EA in the replica  limit 
\bea
\label{eq:replica_limit_vn_asymmetry}
\Delta S_{A}(t)&=&\lim_{n\to 1}\Delta S^{(n)}_{A}(t).
\eea
Given the form~\eqref{eq:symmetrized_RDM_raq} of $\rho_{A, Q}$, the R\'enyi entanglement asymmetry is non-negative~\cite{ASYMM09_sela_insep, hms-23}. As with the case of standard R\'enyi entropies, the higher R\'enyi EAs are not only a convenient calculation tool for determining $\Delta S_A(t)$ but also provide useful information and are accessible experimentally as has been recently shown in~\cite{ASYMM07joshi2024observing}.

The entanglement asymmetry has thus far been studied in several different scenarios, including quench dynamics of  integrable and dissipative models~\cite{ASYMM00asymmetry_og, ASYMM02bertini2023dynamics, ASYMM08rylands2023microscopic, ASYMM01murciano2023xympemba, cma-24, yac-24}, for random circuits \cite{liu_mpemba_circuit-24,tcd-24}, in CFTs~\cite{fadc-24, chch-24}, Haar random states within the context of black hole evaporation~\cite{ampc-23}, confinement and kink dynamics~\cite{kkhpk-23}, and also extended to account for both finite and non-Abelian symmetries~\cite{ASYMM03Ferro_2024,ASYMM04capizzi2023entanglement, ASYMM05capizzi2023universal}.  In the non-equilibrium cases, both the QME as well as other more exotic relaxation has been observed from a variety of initial states.    

\textbf{The Quantum Mpemba Effect:}  We  now return to the original motivating question and consider the quench dynamics of two symmetry broken states, $\rho_1$ and $\rho_2$, evolving under the same Hamiltonian.  The EA provides us with a way to quantify how far the reduced density matrices for these states, $\rho_{A,1}$ and $\rho_{A,2}$,  are from being symmetric and how they approach their stationary (symmetric) values.  Therefore, we have a way to determine if it is possible for a state which is initially further from equilibrium,  i.e.  having more asymmetry, to relax faster.  This leads us to the quantum Mpemba effect which, in terms of the EA, is defined as occurring if initially
\begin{equation}
   \Delta S_{A,1}(t=0)>\Delta S_{A,2}(t=0)
\end{equation}
and there exists a time $t=\tau_{M}$ after the quench when their values reverse and stay reversed for the rest of the evolution
\begin{equation}
    \Delta S_{A,1}(t>\tau_{M})<\Delta S_{A,2}(t>\tau_{M}). 
\end{equation}

In all cases studied in the literature, the crossing signifying the occurrence of the QME happens at times scaling linearly with the subsystem size $\tau_{M}/\ell\sim 1$ and moreover, once the crossing occurs, no subsequent crossings appear.  Thus,  one could be lead to the conclusion that this is generically the case and the existence of the QME could be determined via some short time expansion.  In this work, we address this question and find that this reasoning does not hold.  We show that the crossing of the EA between two states is not restricted to occurring at short times and, moreover, is not restricted to just a single instance.  
In particular, we study the quench dynamics of free fermionic chains, initiated in ground states of free Hamiltonians which have long-range hopping and pairing terms.  We uncover richer relaxation dynamics, in which the EAs of different states can cross multiple times and at late times.  We then interpret this through the quasi-particle picture of entanglement dynamics, thereby finding that this phenomenology can occur in any model which admits such a description.

\textbf{Outline:} The article is organized as follows. In Section \ref{sec:setup}, we introduce the specific quench protocol that we will study and we obtain the ground state of the long-range free fermionic chains that we will use as initial configurations. In Section \ref{sec:dsan_and_charged_moments_peschel}, we describe how the entanglement asymmetry can be calculated through the charged moments of the reduced density matrix and the two-point correlations in the case of fermionic Gaussian states. In Section \ref{sec:renyi_asymmetries_within_qpp}, we derive the time evolution for the charged moments and the entanglement asymmetry using the quasi-particle picture of entanglement. In Section \ref{sec:results}, we present our main results concerning the dynamics of the entanglement asymmetry for quenches from long-range free fermionic chains. In Section \ref{sec:discussion}, we discuss and interpret them in terms of the quasi-particle picture. In Section \ref{sec:Conclusions}, we present our conclusions and discuss future possible continuations of this work.

\section{Quench protocol }
\label{sec:setup}

In this work, we shall study the quench dynamics of the tight-binding model, i.e. time evolution governed by the free fermionic Hamiltonian
\begin{equation}
\label{eq:xx_hamil_real_space}
    H=-\frac{1}{2}\sum_{n=1}^{N}\left(c^{\dagger}_{n}c_{n+1} +c_{n}c^{\dagger}_{n+1}\right),
\end{equation}
where $c^\dag_n,c_n$ are the canonical fermionic creation and annihilation operators on the site $n$ obeying $\{c^\dag_n,c_m\}=\delta_{m,n}$. (This Hamiltonian is equivalent, via a Jordan-Wigner transformation, to the XX spin chain, but the correspondence breaks down for the long-range Hamiltonian corresponding to some initial states and for this reason we will only use fermionic variables rather than spins.)
We assume periodic boundary conditions $c_{n+N}=c_n$.   Introducing the Fourier transform of the fermionic operators, 
\begin{equation}
\label{eq:fourier_space_modes_and_momenta}
b_k=\frac{1}{\sqrt{N}} \sum_{n=1}^N e^{-\frac{i2\pi k n}{N}} c_n, \quad k \in [0,1,\dots, N-1],
%\mathbb{Z},
\end{equation} 
we can rewrite the Hamiltonian in diagonal form,
\be
\label{eq:xx_hamil_in_mom_space_and_disp_rel}
H=\sum_k \epsilon(k) b^\dag_k b_k,\quad \epsilon(k)=-\cos\left(\frac{2\pi k}{N}\right).
\ee
The Fourier modes $b_k^\dag, b_k$ thus describe the quasi-particles of the system,  which propagate with velocity $v(k)\equiv \frac{N}{2\pi}\frac{{\rm d}\epsilon(k)}{{\rm d}k}=\sin(\frac{2\pi k}{N})$.   The particle number
\be
\label{eq:xx_conserved_charge_real_space_mom_space}
Q=\sum_{n=1}^Nc^\dag_n c_n=\sum_{k}b^\dag_k b_k,
\ee 
is one of the conserved charges of the free-fermion Hamiltonian.
It generates a $U(1)$ symmetry with respect to which we shall study the symmetry restoration dynamics emerging from a non-particle number conserving state $\ket{\Psi_0}$.  

In our study, we consider as initial configurations the ground states of a class of free fermionic Hamiltonians,  denoted $H_{0}$, with possible long-range hopping and pairing terms. Following the notation of \cite{ares_long_range_brokensym}, these are given by
\begin{equation}
\label{eq:lr_hop_pair_hamiltonians_general_def}
H_{0}=\frac{1}{2} \sum_{n=1}^N \sum_{l=-R}^R\left(2 A_l c_n^{\dagger} c_{n+l}+B_l c_n^{\dagger} c_{n+l}^{\dagger}-B_l c_n c_{n+l}\right),
\end{equation}
where $R<N/2$ is the maximum range of the hopping and pairing terms.  We assume that the hopping and pairing amplitudes $A_l$ and $B_l$ are real. They must satisfy $A_{-l}=A_l$ and $B_{-l}=-B_l$ for the Hamiltonian $H_0$ to be Hermitian. As long as the pairing terms $B_{l}$ are non zero for some $l$, the Hamiltonian breaks particle number conservation, i.e. $[Q,H_0]\neq 0$, and accordingly its ground state shall also.  In the case in which there are only next-neighbour terms, $R=1$, we recover the fermionic version of the XY spin chain. 
Our setup thus generalizes the study of \cite{ASYMM01murciano2023xympemba}, where the quenches to the XX spin chain from the symmetry breaking ground states of the XY model were considered. The general case of non-local pairing and hopping, which is imprinted on its ground states, shall lead to a richer dynamics of the EA.

\subsection{Ground state of $H_{0}$}
\label{sec:diagonalization_of_LR_Hamiltonians}

To determine the ground state of $H_0$, we write it in terms of the Fourier creation and annihilation operators~\eqref{eq:fourier_space_modes_and_momenta}, 
\begin{equation}
\label{eq:lr_hop_pair_hamiltonians_mom_space}
H_0=\mathcal{E}+\frac{1}{2} \sum_{k=0}^{N-1}\left(b_k^{\dagger}, b_{-k}\right)\left(\begin{array}{cc}
F_k & G_k \\
\bar{G}_k & -F_{-k}
\end{array}\right)\left(\begin{array}{c}
b_k \\
b_{-k}^{\dagger}
\end{array}\right),
\end{equation}
with 
\begin{equation}
\label{eq:lr_hop_pair_hamiltonians_fk_gk_defs}
\begin{aligned}
&F_k=\sum_{l=-R}^R A_l e^{\frac{i 2\pi k l}{N}},\quad 
&G_k=\sum_{l=-R}^R B_l e^{\frac{i 2\pi k l}{N}},\quad 
\mathcal{E}=\frac{1}{2} \sum_{k=0}^{N-1} F_k,
\end{aligned}
\end{equation}
where $F_{k}$ is a real, even function of $k$ and $G_{k}$ an imaginary and odd function of $k$. At this point, we need to make a Bogoliubov rotation to the Hamiltonian written in momentum space to diagonalize it. The matrix 
\begin{equation}
\label{eq:lr_hop_pair_hamiltonian_matrix_mom_space}
M_k=\left(\begin{array}{cc}
F_k & G_k \\
\bar{G}_k & -F_{-k}
\end{array}\right),
\end{equation}
which enters in Eq.~\eqref{eq:lr_hop_pair_hamiltonians_mom_space}, can be diagonalized using a unitary similarity transformation $U_{k}$,
\begin{equation}
\label{eq:lr_hop_pair_hamiltonian_rotated_matrix}
U_k M_k U_k^{\dagger}=\left(\begin{array}{cc}
\Lambda_k & 0 \\
0 & -\Lambda_{-k}
\end{array}\right),
\end{equation}
where $\Lambda_k\geq 0$, with $\Lambda_{k+N}=\Lambda_{k}$, is the dispersion relation of the Bogoliubov modes, whose creation and 
annihilation operators $d_k^\dagger$, $d_{k}$ are given by
%In addition, $R_{k}$ satisfies
%\begin{equation}
%R_{-k}=-\left(\begin{array}{ll}
%0 & 1 \\
%1 & 0
%\end{array}\right) \bar{R}_k\left(\begin{array}{ll}
%0 & 1 \\
%1 & 0
%\end{array}\right)
%\end{equation}
%and to define a specific order of the eigenvalues to find the ground state, we demand $\frac{\Lambda_{k}+\Lambda_{-k}}{2}\geq 0$. 
%Defining the Bogoliubov modes $d_{k}, d^{\dagger}_{-k}$ as 
\begin{equation}
\left(\begin{array}{c}
d_k \\
d_{-k}^{\dagger}
\end{array}\right)=U_k\left(\begin{array}{c}
b_k \\
b_{-k}^{\dagger}
\end{array}\right).
\end{equation}
In terms of them, $H_0$ is diagonal, 
\begin{equation}
H_0=\mathcal{E}+\sum_{k=0}^{N-1} \Lambda_k\left(d_k^{\dagger} d_k-\frac{1}{2}\right). 
\end{equation}
To find $\Lambda_{k}$, we can use the fact that the trace and the determinant of the matrix~\eqref{eq:lr_hop_pair_hamiltonian_matrix_mom_space} are invariant under a change of basis. Taking into account that $F_{k}=F_{-k}$  and $iG_{k}\in\mathbb{R}$, it follows that
\begin{equation}
\Lambda_k=\sqrt{F_k^2+\left|G_k\right|^2}\geq0.
\end{equation}
Since the dispersion relation is non-negative for all the modes, the ground state $\ket{{\rm GS}}$ is the Bogoliubov vacuum, the state annihilated by $d_k$, i.e. $d_k\ket{{\rm GS}}=0$, for all $k$. 

In the rest of the manuscript, we will consider the thermodynamic limit $N\to\infty$, in which the discrete values of the momenta 
$\frac{2\pi k}{N}$ become a continuous variable $\lambda\in[-\pi, \pi)$ (or equivalently $[0,2\pi)$) and the functions $F_k$ and $G_k$ in Eq.~\eqref{eq:lr_hop_pair_hamiltonians_fk_gk_defs} 
\begin{equation}
\label{eq:occup_func_lr_hamilt_gss_continuum}
\begin{aligned}
&F(\lambda)=2\sum_{l=1}^R A_l \cos{(\lambda l)} +A_{0},\quad 
&G(\lambda)=2i\sum_{l=1}^R B_l \sin{(\lambda l)}
\end{aligned}.
\end{equation}

\subsection{Ground state correlation matrix}
\label{sec:corrmatr_of_the_ground_states}
Since we are studying the unitary dynamics of a free fermionic model~\eqref{eq:xx_hamil_real_space} initialized in the ground state of a different free fermionic Hamiltonian~\eqref{eq:lr_hop_pair_hamiltonians_general_def}, the time evolved state $\ket{\Psi(t)}$ satisfies Wick's theorem at any time and all the information about the system is captured by the two-point correlation functions. In momentum space, there are four non trivial correlators, namely $\left<\right.\!b^{\dagger}_{\lambda}b_{\lambda}\!\left.\right>$, $ \left<\right.\!b^{\dagger}_{\lambda}b^{\dagger}_{-\lambda}\!\left.\right>$, and their Hermitian conjugates. 

%Here  the expectation value is evaluated on the ground state of $H_{0}$, which corresponds with the vacuum of the Bogoliubov modes, since $\Lambda_{k}>0, \forall k$.   

The mode occupation function in the initial state, $\vartheta(\lambda)=\bra{{\rm GS}}b^{\dagger}_{\lambda}b_{\lambda}\ket{{\rm GS}}$, is of particular importance for the calculation of the entanglement asymmetry $\Delta S_{A}^{(n)}(t)$.
Taking into account that $\ket{{\rm GS}}$ is the vacuum of the Bogoliubov modes, 
we directly find in the thermodynamic limit that
\begin{equation}
\label{eq:occup_func_lr_hamilt_gss_discrete}
    \vartheta(\lambda)=\frac{1}{2}\left(1+\frac{F(\lambda)}{\sqrt{F(\lambda)^{2}+|G(\lambda)|^{2}}}\right).
\end{equation}
A key point to note, which will be referenced in subsequent sections regarding the multiple crossings of $\Delta S_{A}^{(n)}(t)$, is that $F(\lambda)$ and $G(\lambda)$, defined in Eq.~\eqref{eq:occup_func_lr_hamilt_gss_continuum}, are the sum of oscillating functions of the wave-vectors $\lambda$.
The frequency of oscillation of each term in the sum is controlled by the hopping and pairing range $l=1,...,R$.  This leads us to the conclusion that the momentum occupation function $\vartheta(\lambda)$ for the initial state of our quench can in general have a very complicated structure.  Specifically, it can be an oscillating  function of the momentum $\lambda$, meaning that our initial state would have an oscillating modulation in the population of the momenta.  As a result, the post quench dynamics can have a much richer structure when longer range hopping and pairing terms are included. 

The pair creation and pair annihilation correlation functions $p(\lambda)=\bra{{\rm GS}} b_\lambda^\dagger b_{-\lambda}^\dagger \ket{{\rm GS}}$, $\tilde{p}(\lambda)=\bra{{\rm GS}} b_\lambda b_{-\lambda} \ket{{\rm GS}}$, which are non zero due to the breaking of the particle number symmetry in the ground state, are given in the thermodynamic limit by
\begin{equation}
\label{eq:other_elements_of_corrmatr_result_continuum}
   \tilde{p}(\lambda)= p(\lambda)=\frac{1}{2}\frac{G(\lambda)}{\sqrt{F(\lambda)^{2}+|G(\lambda)|^{2}}}.
\end{equation}

\section{Charged moment decomposition of the R\'enyi entanglement asymmetries and their time evolution}
\label{sec:dsan_and_charged_moments_peschel}

In this section, we review the calculation of $\Delta S^{(n)}_{A}(t)$ in terms of the charged moments of the EA, which in turn can be calculated for free fermionic systems either exactly or using an emergent quasi-particle picture (QPP).
After applying the definition of $S^{(n)}\left(\rho\right)$, the R\'enyi entanglement asymmetry~\eqref{eq:nth_renyi_asymmetry_def} is given by
\begin{equation}
\label{eq:nth_renyi_asymm_log_of_trace_ratio}
    \Delta S^{(n)}_{A}(t)=\frac{1}{1-n}\log\left(\frac{\operatorname{Tr}\left[\rho^{n}_{A,Q}(t)\right]}{\operatorname{Tr}\left[\rho^{n}_{A}(t)\right]}\right).
\end{equation}
$\operatorname{Tr} \left[ \rho_{A,Q}^{n}(t) \right]$ and $\operatorname{Tr}\left[\rho_{A}^{n}(t)\right]$ are then necessary in order to find $\Delta S_{A}^{(n)}(t)$.  
Hence, we can employ the Fourier representation of the projectors $\Pi_q$, 
\begin{equation}
\Pi_{q}=\int_{-\pi}^\pi \frac{{\rm d}\alpha}{2\pi} e^{i\alpha(q-Q_{A})},
\end{equation}
to cast $\rho_{A,Q}(t)$  in the following form
\begin{equation}
\label{eq:roaq_projectors}
\rho_{A, Q}(t)=\int_{-\pi}^{\pi} \frac{\mathrm{d} \alpha}{2 \pi} e^{-i \alpha Q_A} \rho_A(t) e^{i \alpha Q_A}.
\end{equation}
This allows us to express $\operatorname{Tr} \left[\rho_{A,Q}^{n}(t)\right]$ as 
\begin{equation}
\label{eq:trace_roaq_tothe_n_charged_moments}
    \operatorname{Tr}\left[\rho_{A, Q}^n(t)\right]
    %=\int_{-\pi}^\pi \frac{\mathrm{d} \alpha_1 \ldots \mathrm{d} %\alpha_n}{(2 \pi)^n}\operatorname{Tr}\left[\prod_{j=1}^n %\rho_A(t) e^{i \alpha_{j, j+1} Q_A}\right]
    =\int_{-\pi}^\pi \frac{\mathrm{d} \alpha_1 \ldots \mathrm{d} \alpha_n}{(2 \pi)^n} Z_n(\boldsymbol{\alpha},t),
\end{equation}
having introduced the charged moments $Z_{n}(\boldsymbol{\alpha},t)$ of $\rho_A$
\begin{equation}\label{eq:chargedmoment}
Z_n(\boldsymbol{\alpha},t)=\operatorname{Tr}\left[\prod_{j=1}^n \rho_A(t) e^{i \alpha_{j, j+1} Q_A}\right],
\end{equation}
as well as $\alpha_{i,j}=\alpha_{i}-\alpha_{j}$, $\alpha_{n+1}=\alpha_1$ and $\boldsymbol{\alpha}=(\alpha_{1},...,\alpha_{n})$.  
%In a subsequent section we will see another derivation of the charged moments, which amounts in a change of basis from the relative coordinates $\alpha_{ij}=\alpha_{j}-\alpha_{i}$ to $\alpha_{i}$ themselves subject to the constraint that $\sum_{j=1}^{n}\alpha_{j}=0$ \cite{ASYMM08rylands2023microscopic}.\\
Naturally, the charged moments also give us the neutral moments $\operatorname{Tr}\left[\rho_{A}^{n}(t)\right]$, which enter in the asymmetry formula, when we take $\boldsymbol{\alpha}=\boldsymbol{0}$.  Combining this with the definition~\eqref{eq:nth_renyi_asymm_log_of_trace_ratio} of the EA, we arrive at

\begin{equation}\label{eq:dsan_with_zns}
     \Delta S_{A}(t)=\lim_{n\to 1}\frac{1}{1-n}\log\left( \int_{-\pi}^\pi \frac{\mathrm{d} \alpha_1 \ldots \mathrm{d} \alpha_n}{(2 \pi)^n} \frac{Z_n(\boldsymbol{\alpha},t)}{ Z_n(\boldsymbol{0},t)}\right).
\end{equation}
Therefore, the key to evaluating the EA is to calculate the charged moments $Z_n(\boldsymbol{\alpha},t)$. Fortunately, this is indeed possible  in free fermionic systems.  Therein we have a Hamiltonian that is quadratic in the fermionic operators and therefore we can calculate $Z_{n}(\boldsymbol{\alpha},t)$ exactly using the connection between the reduced density matrix and the two-point correlation matrix~\cite{p-03}. This approach is usually employed to calculate different quantities in free fermionic systems and, in particular, it has been applied in \cite{ASYMM01murciano2023xympemba, ASYMM00asymmetry_og,ASYMM06lackofrestoration,ASYMM08rylands2023microscopic} to study the R\'enyi entanglement asymmetry.
%The method is recounted below and used in order to calculate the form of the $n=2$ charged moment.

Since the time evolved state $\ket{\Psi(t)}$ satisfies Wick's theorem,
the reduced density matrix $\rho_A(t)$ is Gaussian, i.e.
\begin{equation}
\label{eq:peschel_RDM_corrmat}
     \rho_{A}=Z^{-1}e^{-\frac{1}{2}\sum_{j,j'}\boldsymbol{c}^{\dagger}_{j}M_{jj'}\boldsymbol{c}_{j'}},%=e^{-\sum_{ij}\boldsymbol{c}^{\dagger}_{i}\log{\frac{I-\Gamma_{ij}}{I+\Gamma_{ij}}\boldsymbol{c}_{j}}}, 
    \quad Z=\Tr(e^{-\frac{1}{2}\sum_{j,j'}\boldsymbol{c}^{\dagger}_{j}M_{jj'}\boldsymbol{c}_{j'}}),
\end{equation}
in terms of the doublet $\boldsymbol{c}_{j}=(c^{\dagger}_{j},c_{j})$. Here $M$ is a $2\ell\times 2\ell$ matrix related to the spatial two-point correlation matrix in $A$,
%are Gaussian, meaning that we can apply Wick's theorem, $\rho_{A}(t)$ is determined directly from the $ 2-$point correlation matrix $\Gamma_{j j^{\prime}}(t)$  \cite{Peschel_2009},  defined as
\begin{equation}
\label{eq:corrmat}
    \Gamma_{j j^{\prime}}(t)=2 \operatorname{Tr}\left[\rho_{A}(t)\boldsymbol{c}_{j}^\dagger\boldsymbol{c}_{j^{\prime}}\right]-\delta_{j j^{\prime}}, \quad j,j^{\prime}\in A,
\end{equation}
via the identity~\cite{p-03}
\begin{equation}
    M(t)=\log\frac{I-\Gamma(t)}{I+\Gamma(t)}.
\end{equation}

Since the charge operator $Q_A$ is a quadratic fermionic operator, then the charged moments are in our case the trace of a product of the Gaussian operators $\rho_A(t)$ and $e^{i\alpha_{j,j+1}Q_A}$. Applying the properties~\cite{balian, Fagotti_2010_disjoint_xy} of the product and trace of this kind of operators,
 we can obtain the following expression for $Z_{n}(\boldsymbol{\alpha},t)$ in terms of the correlation matrix~\cite{ASYMM00asymmetry_og, ASYMM06lackofrestoration},
\begin{equation}\label{eq:exact_nth_Renyichargedmoment}
Z_n(\boldsymbol{\alpha},t)=\sqrt{\operatorname{det}\left[\left(\frac{I-\Gamma(t)}{2}\right)^n\left(I+\prod_{j=1}^n W_j(t)\right)\right]},
\end{equation}
where $W_j(t)=(I+\Gamma(t))(I-\Gamma(t))^{-1} e^{i \alpha_{j, j+1} n_A}$ and $n_{A}$ is a diagonal $2\ell\times 2\ell$ matrix with entries $(n_A)_{2j, 2j}=1$ and $(n_A)_{2j-1, 2j-1}=-1$. This formula allows us to exactly calculate from the knowledge of the correlation matrix the charged moments $Z_{n}(\boldsymbol{\alpha},t)$ and, consequently, the R\'enyi entanglement asymmetries $\Delta S_{A}^{(n)}(t)$.

In the quenches of our interest, given that the system is always translationally invariant, $\Gamma(t)$ is a block Toeplitz matrix, whose elements are given by~\cite{ares_long_range_brokensym}
\begin{equation}
\Gamma_{j j^{\prime}}(t)=\int_{-\pi}^{\pi} \frac{\mathrm{d} \lambda}{2 \pi} \mathcal{G}(\lambda,t) e^{-i \lambda\left(j-j^{\prime}\right)}, \quad j, j^{\prime}=1, \ldots, \ell,
\end{equation}
wherein $\mathcal{G}(\lambda,t)$ is 
\begin{equation}
\label{eq:toeplitz_symbol_ff_quench}
\mathcal{G}(\lambda
, t)=\left(\begin{array}{cc}
1-2 \vartheta(\lambda) & -2 e^{-i 2 t \epsilon(\lambda)} p(\lambda) \\
2 e^{i 2 t \epsilon(\lambda)} p(\lambda) & 2 \vartheta(\lambda)-1
\end{array}\right),
\end{equation}
and  $\vartheta(\lambda)=\bra{{\rm GS}}b_{\lambda}^{\dagger}b_{\lambda}\ket{{\rm GS}}$ and $p(\lambda)=\bra{{\rm GS}}b_{\lambda}^{\dagger}b_{-\lambda}^{\dagger}\ket{{\rm GS}}$ are the two-point correlators in the initial ground state calculated in Eqs.~\eqref{eq:occup_func_lr_hamilt_gss_discrete} and~\eqref{eq:other_elements_of_corrmatr_result_continuum}.

\section{Time evolution of the entanglement asymmetry and the quasi-particle picture}
\label{sec:renyi_asymmetries_within_qpp}

In the preceding section, we obtained in Eq.~\eqref{eq:exact_nth_Renyichargedmoment} a way to calculate  the charged moments $Z_{n}(\boldsymbol{\alpha},t)$ from the correlation matrix for arbitrary time $t$. This is an exact result,  and despite not being written in a transparent analytical form,  it can be calculated efficiently via numerical methods.
In this section, we will derive an analytical prediction for $\Delta S^{(n)}_{A}(t)$ by going to the scaling limit $\ell,t\rightarrow\infty$ keeping $\zeta=t/\ell$ fixed.
In this limit, the system is characterized by the ballistic transport of quasi-particles, which are responsible for the spreading of entanglement and correlations   between the subsystem $A$ and its complement $\bar{A}$~\cite{QUENCH04_Calabrese_2005, QUENCH_05_alba_cal_pnas, QUENCH01_alba_cal_SciPostPhys}.
Specifically, we will deduce the time evolution of the R\'enyi entanglement asymmetry within this QPP obtaining an analytical formula.
Then, we will take the replica limit $n\to 1$ to find $\Delta S_{A}(t)$. 

The basic idea behind the quasi-particle picture is as follows.  At $t=0$, the quench produces sets of correlated quasi-particles  which propagate throughout the system.  For the states which we consider, these are grouped as pairs with equal and opposite momenta~$\{\lambda,-\lambda\}$.   Viewed semi-classically, a pair emerges from a single point in space and is uncorrelated with any other pairs, and the only correlations in the system are between quasi-particles from the same pair.  As the quasi-particles propagate through  the system, with velocity $v(\lambda)$, they carry their correlations with them and thereby create spatial correlation and entanglement between different regions.   Thus, the dynamics of a subsystem observable is dominated at short times by the pairs of particles which have both members completely within the subsystem, we call such pairs complete. On the other hand, at long times,  it is not possible for both members of a pair to be within the subsystem. Therefore, the dynamics is dominated (See Figure~\ref{fig:quasiparticle}) by pairs which have one member in $A$ and the other in $\bar A$. These pairs are referred to as incomplete.

To obtain the quasi-particle prediction 
for the EA, it is necessary to first do so for the charged moments.  The strategy will be to calculate these at $t=0$, thereby understanding the initial contribution of the quasi-particles from complete pairs  and then to look at the long time behaviour which will give us the contribution of a single quasi-particle from an incomplete pair.

\subsection{Initial state charged moments}
\label{sec:initial_state_charged_moments}

We begin by calculating $Z_{n}(\boldsymbol{\alpha},t=0)$ in the limit $\ell\to\infty$ starting from~\eqref{eq:exact_nth_Renyichargedmoment}.  
It can be observed that the matrix $W_{j}(t)$ contains the inverse of  $I-\Gamma$, the inverse of a Toeplitz matrix, that is in general not block-Toeplitz. However, we can apply the following result derived in \cite{ASYMM06lackofrestoration}. If $T_{\ell}[g_{j}] $ is a $d\ell \times d\ell$ block-Toeplitz matrix with symbol $g_{j}$, of dimension $d\times d$, and $T_{\ell}[g^{\prime}_{j}]^{-1} $ the inverse of another one of symbol $g_j'$, then
\begin{equation}
\label{eq:inverse_of_toeplitz_formula}
    \det \left( I+\prod_{j=1}^{n} T_{\ell}[g_{j}] T_{\ell}[g^{\prime}_{j}]^{-1} \right)\propto e^{\ell A^{\prime}_{n}}.
\end{equation}
The exponent $A^{\prime}_{n}$ is given by 
 \begin{equation}
 \label{eq:an_for_inverse_of_toeplitz}
     A^{\prime}_{n}=\int_{-\pi}^{\pi}\frac{\mathrm{d}\lambda}{2\pi}\log \det \left[I+\prod_{j=1}^{n}g_{j}(\lambda)g_{j}^{\prime}(\lambda)^{-1}\right].
 \end{equation}
In our case, however, we cannot apply this result directly since the symbol of $I-\Gamma$ is $I-\mathcal{G}(\lambda)$, which is non-invertible.
This problem is remedied by considering that our system is at finite temperature $1/\beta$, allowing for the inversion of the correlation matrix symbol, and then taking the zero temperature limit, $\beta\rightarrow\infty$~\cite{ASYMM01murciano2023xympemba}.
The finite temperature state is described by the Gibbs ensemble of $\rho_{\beta}=e^{-\beta H_0}/Z$, with the normalization constant $Z=\operatorname{Tr}(e^{-\beta H_0})$.
For the subsystem $A$, this leads to a temperature dependent Gaussian reduced density matrix $\rho_{A,\beta}$. The corresponding correlation matrix $\Gamma_{\beta}$ is block Toeplitz with symbol
\begin{equation}
\label{eq:temperature_dep_toeplitz_symbol}
    \mathcal{G}_{\beta}(\lambda)=\tanh{\left(\frac{\beta\Lambda_{\lambda}}{2}\right)}\left(\begin{array}{cc}
1-2 \vartheta(\lambda) & -2 e^{-i 2 t \epsilon(\lambda)} p(\lambda) \\
2 e^{i 2 t \epsilon(\lambda)} p(\lambda) & 2 \vartheta(\lambda)-1
\end{array}\right).
\end{equation}
Since now $I-\mathcal{G}_{\beta}(\lambda)$ is invertible, we can apply~\eqref{eq:inverse_of_toeplitz_formula} and after taking the $\beta\rightarrow\infty$ we can get the $n-$th charged moment  at $t=0$,
\begin{equation}
\label{eq:zero_temp_lim_of_cm}
    Z_{n}(\boldsymbol{\alpha},0)=\lim_{\beta\to \infty}e^{\ell A_{n}^{\prime}(\boldsymbol{\alpha},\beta)},
\end{equation}
with  $A_{n}^{\prime}(\boldsymbol{\alpha},\beta)$ given by
\begin{equation}
\label{eq:an_for_cm_with_corrmatr_explicitely}
    A_{n}^{\prime}(\boldsymbol{\alpha},\beta)=\int_{-\pi}^{\pi}\frac{\mathrm{d}\lambda}{2\pi}\log\operatorname{det}\left[\left(\frac{I-\mathcal{G}_{\beta}(\lambda)(t)}{2}\right)^n\left(I+\prod_{j=1}^n \mathcal{W}_{\beta,j}(\lambda)\right)\right],
\end{equation}
wherein the $\beta$ dependent $\mathcal{W}_{j,\beta}(\lambda)=(I+\mathcal{G}_{\beta}(\lambda))(I-\mathcal{G}_{\beta}(\lambda))^{-1} e^{i \alpha_{j, j+1} n_{A}}$. 

Obtaining $\Delta S^{(n)}_{A}(t=0)$ requires to take the $\beta\rightarrow\infty$ limit of the charged moment ratio $\frac{Z_{n}(\boldsymbol{\alpha},\beta)}{Z_{n}(\boldsymbol{0},\beta)}$, that leads to 
\be 
\label{eq:ratio_of_cm_for_asymm_zero_temp}
Z_{n}(\boldsymbol{\alpha}, 0)=Z_{n}(\boldsymbol{0}, 0)e^{\ell A_n(\boldsymbol{\alpha})},
\ee
having defined $A_{n}(\boldsymbol{\alpha})$ for convenience as
\be 
\label{eq:zero_temp_an_for_cm}
A_{n}(\boldsymbol{\alpha})\equiv \lim_{\beta\rightarrow\infty}\left[ A^{\prime}_{n}(\boldsymbol{\alpha},\beta)- A^{\prime}_{n}(\boldsymbol{0},\beta)\right].
\ee

To complete the calculation, we determine the exponent explicitly for several integer  values of $n$ from which we can note that in general it has the following form  
\be\label{eq:initialstatechargedmoment}
A_n(\boldsymbol{\alpha})=\int_{-\pi}^{\pi}\mathrm{d} \lambda\sum_{j=1}^n f_{\alpha_{j,j+1}}\left(\lambda\right),
\ee
with $ f_{\alpha}\left(\lambda\right)$ being
\begin{eqnarray}
\label{eq:falpha_for_ff_quench}
 f_{\alpha}\left(\lambda\right)=\frac{1}{4\pi}\log\left[1-\vartheta(\lambda)+\vartheta(\lambda)e^{2i\alpha}\right]. 
\end{eqnarray}

Interestingly, this function is the same as the one which appears when calculating the full counting statistics of the subsystem charge $Q_A$~\cite{ASYMM02bertini2023dynamics, horvath2023full},  i.e.
\be 
\label{eq:fcs_initial_state_def}
\operatorname{Tr}\left[\rho_A(0)e^{i\alpha Q_A}\right]=e^{\ell \int {\rm d}\lambda \,f_{\alpha}(\lambda)}. 
\ee
This apparent factorization in the initial state into a product of the full counting statistics in each replica is a particular property of the choice of the initial state, for example it can be shown that it does not hold when considering mixed initial states~\cite{ares_murciano_unpublished}.

\subsection{Steady state charged moments}
\label{sec:long_time_charged_moments}
After studying the charged moments in the initial state $Z_{n}(\boldsymbol{\alpha},t=0)$, we go to their long time behavior, $Z_{n}(\boldsymbol{\alpha},t\rightarrow\infty)$.
%Its expression is the same as before, containing a product of block-Toeplitz and inverse of block-Toeplitz matrices.
At late times after the quench, the time dependent entries of the symbol~\eqref{eq:toeplitz_symbol_ff_quench} of $\Gamma(t)$ are rapidly oscillating and drop out when averaged over time, giving us the symbol
\begin{equation}
\label{eq:time_avged_toeplitz_symbol}
\mathcal{G}(\lambda
, t\rightarrow \infty )=\left(\begin{array}{cc}
1-2 \vartheta(\lambda) & 0 \\
0 & 2 \vartheta(\lambda)-1
\end{array}\right).
\end{equation}
Evidently, the two correlators corresponding to pair creation, $\bra{\Psi(t)}c^{\dagger}_{j}c^{\dagger}_{j'}\ket{\Psi(t)}$, and pair annihilation, $\bra{\Psi(t)}c_{j}c_{j'}\ket{\Psi(t)}$, are the ones that go to zero, which suggests the restoration of the particle conservation symmetry at large times.

Contrary to the $t=0$ case, the symbol $I-\mathcal{G}(k,t\rightarrow\infty)$ is invertible, wherein~\eqref{eq:inverse_of_toeplitz_formula} can be applied leading to
\begin{equation}
\label{eq:log_cm_inf_time_after_toeplitz}
    \log Z_{n}(\boldsymbol{\alpha},t\rightarrow\infty)\propto \frac{\ell}{2}\int_{-\pi}^{\pi}\frac{\mathrm{d}\lambda}{2\pi}\log\operatorname{det}\left[\left(\frac{I-\mathcal{G}(\lambda,t\rightarrow\infty)(t)}{2}\right)^n\left(I+\prod_{j=1}^n \mathcal{W}_{j}(\lambda)\right)\right],
\end{equation}
having defined $\mathcal{W}_j(\lambda)=(I+\mathcal{G}(\lambda,t\rightarrow\infty))(I-\mathcal{G}(\lambda,t\rightarrow\infty))^{-1} e^{i \alpha_{j, j+1}n_A}$.
Plugging the time-averaged symbol~\eqref{eq:time_avged_toeplitz_symbol} into~\eqref{eq:log_cm_inf_time_after_toeplitz}, we find
\begin{equation}
\label{eq:large_time_cm}
    \log Z_{n}(\boldsymbol{\alpha},t\rightarrow\infty)\propto \ell \int_{-\pi}^{\pi}{\rm d}\lambda\,h_{n}(\vartheta(\lambda)),
\end{equation}
where the function $h_{n}(x)$ is defined as
\begin{equation}
\label{eq:hn_func_of_ff}
    h_{n}(x)=\frac{1}{2\pi}\log\left[x^{n}+(1-x)^{n}\right]. 
\end{equation}
%Symmetry restoration in the long-time limit is indicated by the fact that  $\log Z_{n}(\boldsymbol{\alpha},t\rightarrow\infty)\propto \log Z_{n}(\boldsymbol{0},t\rightarrow\infty)$. 
This result implies that $ Z_{n}(\boldsymbol{\alpha},t\rightarrow\infty)\sim Z_{n}(\boldsymbol{0},t\rightarrow\infty)$, which indicates the symmetry restoration in the long-time limit. We note also that $h_n(x)$ is the same function that appears when calculating the R\'enyi entanglement entropy. 

\subsection{Quasi-particle picture for the charged moments}
\label{sec:finite_time_qpp_charged_moms}

\begin{figure}[t]
    \centering
    \begin{tikzpicture}[scale=2.5] % Adjust the scale factor as needed
        % Draw the 1D quantum system line
        % \draw[thick] (0,0) -- (6,0) node[right] {$\mathcal{l}$}; % Label for horizontal axis
        \node at (0,-0.1) {0}; % Tick mark at the origin
        
        % Label for vertical axis
        \node at (-0.2,1.3) {$t$};
         \node at (6,-0.18) {$x$};        
       
        % Define regions
        \filldraw[fill=blue!10] (0,0) rectangle (2,1.5); % Adjusted dimensions for the first region
        \node at (1,-0.2) {$\Bar{A}$};
        
        \filldraw[fill=yellow!20] (2,0) rectangle (4,1.5);   % Adjusted dimensions for the second region (light shade of yellow)
        \node at (3,-0.2) {${A}$};
        \filldraw[fill=blue!10] (4,0) rectangle (6,1.5); % Adjusted dimensions for the third region
        \node at (5,-0.2) {$\bar{A}$};
        \node at (3.14,1.2) {$k$};
        \node at (2.81,1.2) {$-k$};
        % \node at (2.25,1.2) {$v(k)$};
        % \node at (2.4,0.42) {$v(k)$};
        \draw[-latex] (0,0) -- (0,1.6) ; % Y-axis
        \draw[-latex] (0,0) -- (6.1,0) ; % Y-axis
        % Draw particle trajectories with dots at the end
        \foreach [count=\i] \x in {1.2, 3.85, 2.97}
        {
            \ifnum\i=3 
                \draw[-latex,blue,very thick] (\x,0) -- ++(75:1.3) node[circle,fill,inner sep=1pt] {}; % Extended length for all trajectories with dot at the end
                \draw[-latex,blue,very thick] (\x,0) -- ++(105:1.3) node[circle,fill,inner sep=1pt] {}; % Extended length for all trajectories with dot at the end
            \fi    
            \ifnum \i=2 
% \draw[dashed] (200pt,40pt) -- (60pt,40pt);
% \draw (60pt,40pt) -- (0pt,40pt);
                %\draw[red,very thick](\x,0) -- ++(75:0.56);
                \draw[-latex,red,dotted,very thick](\x,0) -- ++(75:1.3) node[circle,fill,inner sep=1pt] {}; 
                \draw[red,very thick] (\x,0) -- ++(75:0.58) ; % Extended length for all trajectories with dot at the end
                \draw[-latex,red, very thick] (\x,0) -- ++(105:1.3) node[circle,fill,inner sep=1pt] {}; % Extended length for all trajectories with dot at the end
            \fi
            \ifnum \i=1
% \draw[dashed] (200pt,40pt) -- (60pt,40pt);
% \draw (60pt,40pt) -- (0pt,40pt);
                \draw[-latex,red,dotted,very thick] (\x,0) -- ++(75:1.3) node[circle,fill,inner sep=1pt] {}; % Extended length for all trajectories with dot at the end
                \draw[-latex,red,dotted, very thick] (\x,0) -- ++(105:1.3) node[circle,fill,inner sep=1pt] {}; % Extended length for all trajectories with dot at the end
            \fi            
        }
        % Additional point emitting trajectories into the middle partition with dots at the end
        % \draw[fill=black] (3,0) circle (0.03); % Smaller circle
        % \draw[-latex,red] (3,0) -- ++(45:0.8) node[circle,fill,inner sep=1pt] {}; % Extended length for the right-going trajectory with dot at the end
        % \draw[-latex,red] (3,0) -- ++(135:0.8) node[circle,fill,inner sep=1pt] {}; % Extended length for the left-going trajectory with dot at the end
    \end{tikzpicture}
    \caption{Quasi-particle picture of entanglement asymmetry. After the quench, pairs of entangled quasi-particles produced at the same point propagate ballistically with opposite momentum $\pm k$ and velocity $v(\pm k)$. At time $t$, only those pairs with both excitations inside the subsystem $A$ (complete pairs, solid blue lines) contribute to the entanglement asymmetry~\eqref{eq:QPP_renyi_EA}. Once one of the two excitations leave the subsystem, the pair does not contribute anymore (incomplete pair, red lines). At time $t\to\infty$, there are no complete pairs and the symmetry is restored in $A$.}
     \label{fig:quasiparticle}
\end{figure}
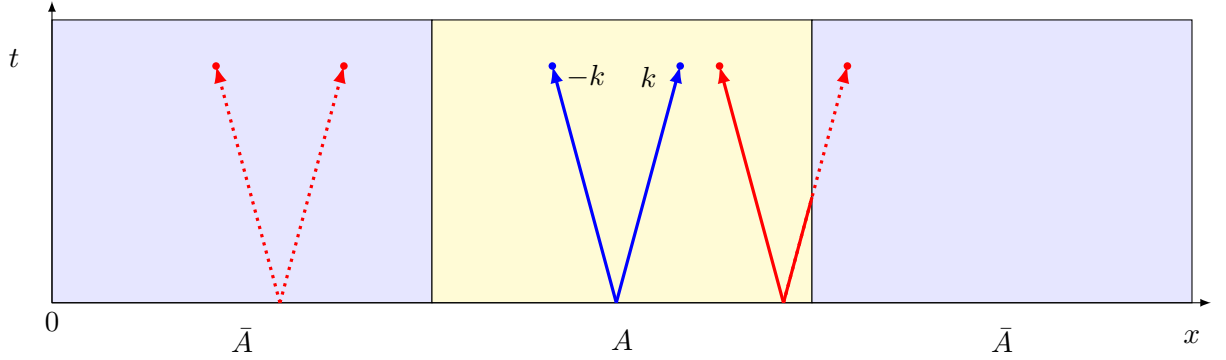

Once we have obtained the expressions for the charged moments at both long times and in the initial state, we may identify the contributions of the  complete and incomplete quasi-particle pairs and construct their dynamics.  To this end, we recall that using simple geometric arguments, see Figure~\ref{fig:quasiparticle}, the number of  quasi-particle pairs of momentum $\{\lambda,-\lambda\}$ forming complete pairs at time $t$ is
\be
n_{\rm c}(t, \lambda)=\ell-{\rm{min}}[2|v(\lambda)| t,\ell]. 
\ee
Likewise, the number of quasi-particle pairs of momentum $\{\lambda,-\lambda\}$ with only one member inside $A$, i.e. the number of incomplete pairs, at time $t$ is
\be
n_{\rm i}(t, \lambda)={\rm{min}}[2|v(\lambda)| t,\ell], 
\ee
which can be determined from particle number conservation which fixes  $n_{\rm c }+n_{\rm i}=\ell$. Furthermore, using the calculation in Eq.~\eqref{eq:initialstatechargedmoment} of the charged moments in the initial state, we can identify the leading order contribution to the charged moments due to a complete pair of momentum $\{\lambda,-\lambda\}$ as being $\sum_{j}f_{\alpha_{j,j+1}}(\lambda)$.  From the long time calculation~\eqref{eq:large_time_cm}, we can instead infer that the  contribution of a quasi-particle from an incomplete pair is given by $h_n(\vartheta(\lambda))/2\pi$.  

Combining them, we find 
\bea \label{eq:chargedmomentqpp}
Z_{n}(\boldsymbol{\alpha},t)&=&Z_n(\boldsymbol{0},0)\exp \left[\int_{-\pi}^\pi {\rm d}\lambda\,  n_{\rm i}(\lambda,t) h_n(\lambda)+n_{\rm c}(\lambda,t)\sum_{j=1}^{n}f_{\alpha_{j,j+1}}(\lambda)\right]\\
&=&Z_n(\mathbf{0}, t)
\exp \left[\ell \sum_{j=1}^n \int_{-\pi}^\pi {\rm{d}} \lambda\,  x_\zeta(\lambda) f_{\alpha_{j,j+1}}(\lambda)\right],
\eea
where we have introduced
\begin{equation}
    x_{\zeta}(\lambda)={\rm max}\left[1-\frac{2t}{\ell}|v(\lambda)|,0\right],
\end{equation}
and we have extracted the contribution of the part independent of $\alpha_{j, j+1}$,  which corresponds to the neutral moments $Z_n(\boldsymbol{0},t)$  of $\rho_A(t)$ .

At this point, we can now return to the EA and, using in Eq.~\eqref{eq:dsan_with_zns} our expression for the charged moments~\eqref{eq:chargedmomentqpp}, we obtain
\bea\label{eq:QPP_renyi_EA}
\Delta S^{(n)}_A(t)&=&\frac{1}{1-n}\log \int_{-\pi}^\pi \frac{{\rm d}\boldsymbol{\alpha}}{(2\pi)^n}\exp \left[\ell \sum_{j=1}^n \int_{-\pi}^\pi {\rm{d}} \lambda\,  x_\zeta(\lambda) f_{\alpha_{j,j+1}}(\lambda)\right]. 
\eea
From this result, we note that only the complete quasi-particle pairs contribute to the EA. This can be contrasted with other  quantities like the entanglement entropy~\cite{QUENCH04_Calabrese_2005, QUENCH_05_alba_cal_pnas, QUENCH01_alba_cal_SciPostPhys,c-18}, which receives contributions only from pairs of quasi-particles which are shared between regions, i.e. incomplete pairs, or the full counting statistics, which has contributions from both types~\cite{ASYMM02bertini2023dynamics, horvath2023full}. In the present case, the contribution of the incomplete pairs has been removed by taking the ratio with the neutral charge moment in~\eqref{eq:dsan_with_zns}. The picture is however consistent, as pairs propagate through the system, they spread correlations to different regions, so a pair shared between $A$ and $\bar A$ contributes to entanglement. On the other hand, the particle number symmetry breaking is a result of correlated pairs within a region and, as entanglement is spread, the symmetry is restored locally.  

The expression for the charged moments using the quasi-particle picture can be derived explicitly applying the exact result of Eq.~\eqref{eq:exact_nth_Renyichargedmoment} for $n=2$ and the multi-dimensional stationary phase approximation~\cite{QUENCH00_calabrese_xy}.  Carrying this out however becomes increasingly more complicated at higher R\'enyi index $n$, but instead the two can be compared after numerical evaluation.  We perform these checks below for specific cases. 

\subsection{The replica limit}
\label{sec:vN_asymmetry_replica_limit}

The main advantage of using the  expression~\eqref{eq:chargedmomentqpp} over the exact result~\eqref{eq:exact_nth_Renyichargedmoment} is that the former  affords a level of analytic control which we can use to take the replica limit $n\to 1$ and recover $\Delta S_A(t)$.  To carry this out, we first make a change of variables $\alpha_{j,j+1}\mapsto \alpha_j'$, with $\alpha_n'=-\sum_{j=1}^{n-1}\alpha_j'$,
\bea\label{eq:resum_renyi_EA}
\Delta S^{(n)}_A(t)&=&\frac{1}{1-n}\log \int_{-\pi}^\pi \frac{{\rm d}\boldsymbol{\alpha'}}{(2\pi)^{n-1}}\exp \left[\ell \sum_{j=1}^n \int_{-\pi}^\pi {\rm{d}} \lambda\,  x_\zeta(\lambda) f_{\alpha'_{j}}(\lambda)\right]\delta_{p}\Big (\sum_{j=1}^n\alpha'_j\Big ). 
\eea
where $\delta_{p}(x)=\frac{1}{2\pi}\sum_{k\in\mathbb{Z}} e^{-i k x}$ is the $2\pi$-periodic delta function which implements the constraint on $\alpha'_n$. We can then reexpress~\eqref{eq:resum_renyi_EA} as 
\be \label{eq:Renyi_with_J}
\Delta S^{(n)}_A(t)=\frac{1}{1-n}\log \sum_{k \in \mathbb{Z}} [J_k(t)]^n,
\ee
where $J_k(t)$ is a single replica quantity given by
\be\label{eq:Jk}
J_k(t)=\int_{-\pi}^\pi \frac{{\rm d}\alpha}{2\pi}e^{-ik\alpha+\ell \int_{-\pi}^\pi {\rm{d}} \lambda\,  x_\zeta(\lambda) f_{\alpha}(\lambda)}. 
\ee
This has the property that $\sum_{k\in\mathbb{Z}}J_k(t)=1$, which can be confirmed by simply using the definition of $\delta_p(x)$ and that $f_0(\lambda)=0$. In addition, we have that $J_{2k-1}(t)=0$ and $J_k(t)=J^*_k(t)$ \cite{ASYMM08rylands2023microscopic}. 

We are now in a position to take the replica limit.  We analytically continue Eq.~\eqref{eq:Renyi_with_J} from $n\in\mathbb{N}$ to $z\in\mathbb{C}$ and use a minimal choice for $J^{z}_{k}$, which drops to the correct result for $z\in\mathbb{N}$. Since the resulting function has to be real, we take
\begin{equation}
     [J_{k}(t)]^n\rightarrow\frac{1}{2}\left(e^{z \log J_{k}}+e^{z \log J^{*}_{k}}\right).
\end{equation}
Inserting it into~\eqref{eq:Renyi_with_J} and taking the replica limit, we obtain
\begin{equation}
\label{eq:dsa_jk_sum_qpp_final}
    \Delta S_{A}(t)=-\sum_{k=-\infty}^{\infty}\Re\left[J_{k}(t)\log J_{k}(t)\right],
\end{equation}
which gives us the full time dynamics of the von Neumann EA.   This expression is particularly convenient,  not only for performing calculations but also for developing a phenomenological understanding of the EA.  Indeed it was the starting point for deriving the set of conditions for which the QME occurs~\cite{ASYMM08rylands2023microscopic}.

In the initial state,  $J_k(0)$ can be shown to be the same as the full counting statistics of the subsystem charge and so $\Delta S_A(t)$ is simply the classical Shannon entropy of the charge probability distribution.  At finite time,  $J_k(t)$ no longer has this interpretation since the dynamics of the full counting statistics is different.  Indeed, as mentioned above, both complete and incomplete pairs contribute to the full counting statistics. In the present case, it can be shown that~\cite{horvath2023full}
\begin{eqnarray}\label{eq:FCS}
\log\left( \operatorname{Tr} \left[\rho_A(t)e^{i\alpha Q_A}\right]\right)= \int {\rm d}\lambda \left[n_{\rm c}(\lambda,t)f_\alpha(\lambda)+2n_{\rm i}(\lambda,t)f_{\alpha/2}(\lambda)\right].
\end{eqnarray}
Here, we recognize the first and second terms as the contributions of the complete and incomplete pairs respectively.  Thus, we can interpret $J_k(t)$ as being the Fourier transform of the contribution of complete pairs to the full counting statistics.

\section{Results}
\label{sec:results}

We now arrive at the main part of our analysis and present our results for the dynamics of the EA in quenches of free-fermion chain~\eqref{eq:xx_hamil_real_space} from the ground state of long-range Hamiltonians of the form~\eqref{eq:lr_hop_pair_hamiltonians_general_def}. 
We choose two specific cases for $H_0$, one in which there are only hoppings between nearest neighbour sites but for which pairing occurs beyond nearest neighbours and a second in which both types of terms are long range.  For context, we also briefly re-examine the case in which both the hopping and pairing are nearest neighbour. 

Our ultimate interest is in the von Neumann entanglement asymmetry.  However, as discussed in the previous sections, to study this we must employ the replica trick and then also invoke a quasi-particle picture to take the replica limit.  To check the validity of this approach, we compare the results of numerically evaluating the exact expression~\eqref{eq:exact_nth_Renyichargedmoment} with the quasi-particle prediction~\eqref{eq:QPP_renyi_EA}
 for the R\'enyi EA.  We present only the results for $\Delta S_A^{(2)}(t)$ and find excellent agreement in the scaling limit of large $t,\ell$.  We have checked that this holds also for integer $n>2$ as well as for a range of $t,\ell$ although for ease of presentation we include only the plots for $n=2$. 
 
After performing this cross check on the quasi-particle picture, we show the results for the EA obtained  using Eqs.~\eqref{eq:dsa_jk_sum_qpp_final} and~\eqref{eq:Jk}.  For the former, we must truncate the sum over integers, $\sum_{k\in\mathbb{Z}}\to \sum_{k=-c_l}^{c_u}$, with $c_{l,u}>0$. The values of  the upper and lower cutoffs are fixed so that the sum rule $\sum J_k(t)=1$ is satisfied to a desired tolerance.  In our scheme, these are allowed to be time dependent for better performance. Furthermore, we also check at regular intervals that  $J_{2k-1}(t)=0$ and $J_k(t)=J^*_k(t)$ are also satisfied.

\subsection{Quench from the short-range pairing Hamiltonian}
\label{sec:NN_hopping_and_NN_pairing_Hamilts}

\begin{figure}[t]
    \centering
   \includegraphics[width=.49\columnwidth]{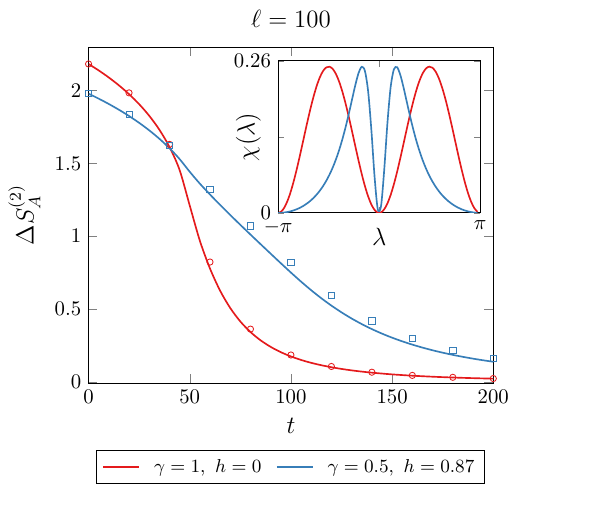} 
    \includegraphics[width=.49\columnwidth]{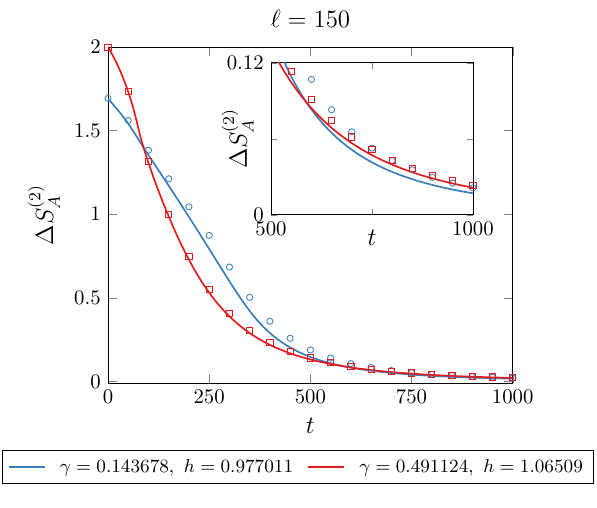} 
\caption{R\'enyi entanglement asymmetry $\Delta S_A^{(2)}$ after a quench to the free-fermion chain from the ground state of the pairing model chain~\eqref{eq:xy_ham} with different couplings $h$ and $\gamma$ and subsystem sizes $\ell=100$ (left) and $150$ (right). In the left panel, we consider two initial states whose EAs intersect at a finite time and the symmetry is restored faster in the more asymmetric initial state than in the less asymmetric one, indicating the occurrence of the QME. In the inset, we represent the mode occupation of Cooper pairs in the two initial configurations. In the right panel, we take two other initial configurations for which their EAs intersect twice and, therefore, the symmetry is eventually restored earlier in the quench from the less asymmetric configuration. The inset is a zoom of the second crossing. In both panels, the symbols are the exact numerical value obtained using Eq.~\eqref{eq:exact_nth_Renyichargedmoment} and the curves correspond to the quasi-particle picture prediction~\eqref{eq:QPP_renyi_EA}.}
\label{fig:xy}
 \end{figure}

 We start by re-examining the case where the initial Hamiltonian~\eqref{eq:lr_hop_pair_hamiltonians_general_def} has only short-range hopping and pairing, 
\begin{equation}
\label{eq:xy_ham}
H_{0}=-\frac{1}{2}\sum_{n=1}^N\left[2h a_{n}^{\dagger}a_{n} +(a_n^{\dagger}a_{n+1}+a_{n+1}^{\dagger}a_n) +\gamma\left( a_n^{\dagger} a_{n+1}^{\dagger}-a_n a_{n+1}\right) \right],
\end{equation}
which is the Hamiltonian of the XY spin chain after a Jordan-Wigner transformation. It explicitly breaks the particle number symmetry when the pairing amplitude $\gamma$ is not zero. The dynamics of the EA evolving from the ground states of~\eqref{eq:xy_ham} were extensively studied in~\cite{ASYMM01murciano2023xympemba}.  Therein, numerous examples of the occurrence of QME were found for specific choices of pairs of ground states for different values of the parameters $h$ and $\gamma$.  We report as an example a typical case in the left panel of Figure~\ref{fig:xy}.  Here we see that the entanglement asymmetry starts at a non-zero value for both states and decays monotonically to zero.  Moreover, we see that for this choice the QME effect occurs.  The curves cross at a specific time, $\tau_{\rm M}$, which is of the order $\tau_M/\ell\sim 1$ and do not cross again for any times thereafter.  As discussed in the introduction, a qualitatively similar picture has been seen in all such examples of the QME observed so far.  In the right panel of Figure~\ref{fig:xy}, we present the results for a different pair of initial states which lie close to the critical line of $H_0$, $h=\pm 1,~\gamma\neq 0$, and show a different behaviour. The state depicted in red lies below the critical line $h=1$, while the state depicted in blue lies above $h=1$ and has larger pairing coefficient.
In this example, we see that once again the states exhibit a crossing at times $t/\ell\sim 1$ but, after a certain time, they cross again.  Thus, in this case, the QME is avoided and the initially more symmetric state relaxes faster. 

This ``fake'' quantum Mpemba effect is a quite fine tuned example and indeed as shown in~\cite{ASYMM01murciano2023xympemba} the vast majority of the pairs of states in the parameter space $(h, \gamma)$ either show a true QME or their EAs never intersect.  In the next sections, we show more robust examples wherein this  avoided QME occurs.

\subsection{Quench from the long-range Kitaev chain}
\label{sec:lr_and_sr_hamiltonians}
We now move on to the quench of ground states of long-range free fermionic Hamiltonians, exhibiting a double crossing of $\Delta S_{A}(t)$. We select two different families of Hamiltonians and find their ground states for two sets of couplings.
Then, we study the symmetry breaking evolution under the quench to the free-fermion chain~\eqref{eq:xx_hamil_real_space} with the methods discussed above.

\subsubsection{Nearest neighbour hopping with power-law pairing}
\label{sec:NN_hopping_and_lr_pairing_hamilts}

\begin{figure}[t]
    \centering
    \includegraphics[width=0.99\textwidth]{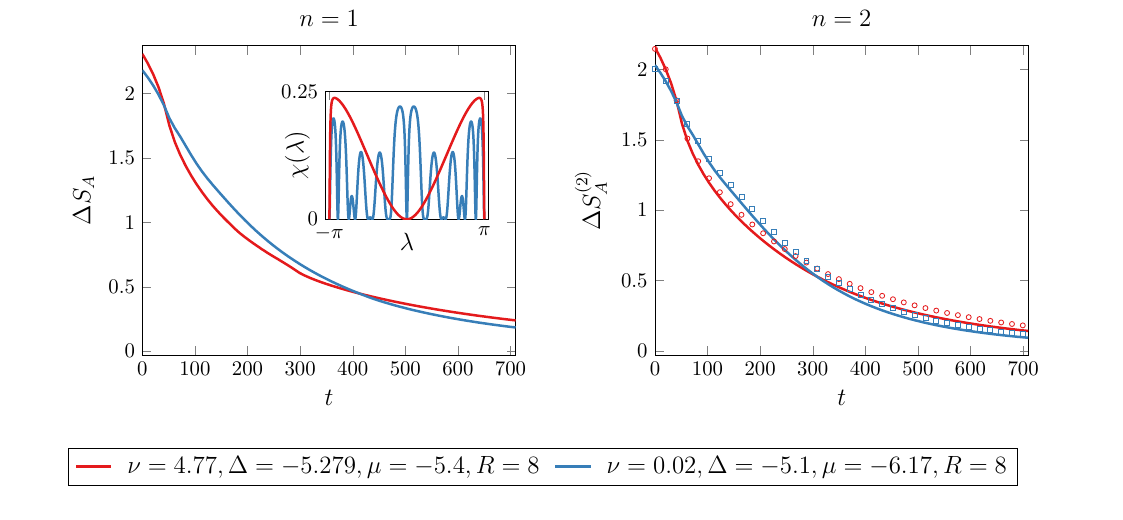} 
    \caption{ Left panel: time evolution of $\Delta S_A(t)$  obtained using Eq.~\eqref{eq:dsa_jk_sum_qpp_final} after the quench to the free-fermion short-range chain from the ground state of~\eqref{eq:long_range_Kitaev} for two different sets of parameters with range $R=8$ for both states. In the inset, we represent the mode occupation of Cooper pairs in the two initial configurations. Right panel: time evolution of $\Delta S^{(2)}_A(t)$ for the same initial states as in the left panel. The symbols are the exact value of the R\'enyi EA calculated numerically with Eq.~\eqref{eq:exact_nth_Renyichargedmoment} while the continuous curves correspond to the quasi-particle prediction~\eqref{eq:QPP_renyi_EA}. In both panels, we take a subsystem of length $\ell=100$. We see that the avoidance of the QME signified by the double crossing of the entanglement asymmetry curves.}
    \label{fig:dsa_qpp_double_crossing}
\end{figure}

The first class of Hamiltonians we consider has nearest-neighbour hopping terms and long-range pairing terms with a power-law form  similar to the ones introduced in~\cite{vodola_prl_2014}, see also~\cite{ares_lr_kitaev}. Models of this type are typically referred to as long-range Kitaev models.  Their Hamiltonian takes the  form
\begin{equation}\label{eq:long_range_Kitaev}
H_{0}=\sum_{n=1}^N\left[-\mu a_{n}^{\dagger}a_{n} -{\frac{\Delta}{2}} (a_n^{\dagger}a_{n+1}+a_{n+1}^{\dagger}a_n) + \frac{\Delta}{2}\sum_{l=-R}^R|l|^{-\nu-1}l\left( a_n^{\dagger} a_{n+l}^{\dagger}-a_n a_{n+l}\right)\right],
\end{equation}
with chemical potential $\mu$ and hopping amplitude $\Delta/2$. The exponent $\nu>0$ tunes the damping of the pairing strength with the distance. The pairing is truncated at a range $R$.

In Figure~\ref{fig:dsa_qpp_double_crossing}, we present plots of the von Neumann (left panel) and the second R\'enyi entanglement asymmetry (right panel) as a function of time for two choices of the parameters $R$, $\mu$, $\nu$, and $ \Delta$. The curves and symbols in red correspond to the ground state (state 1) of a Hamiltonian with a much larger damping exponent $\nu$ than the ones in blue (state 2). In the right panel, we plot $\Delta S_A^{(2)}(t)$ for the two states obtained using the exact expression~\eqref{eq:exact_nth_Renyichargedmoment} (symbols) and using the quasi-particle prediction of~\eqref{eq:QPP_renyi_EA} (continuous lines). We see that the two approaches have good agreement and exhibit the same behaviour: while state 2 has an initially higher value of $\Delta S^{(2)}_A(t)$ than state 1, the two quickly reverse order; however, a subsequent crossing occurs at a later time and the initial ordering is reinstated. We have checked that the discrepancy between the quasi-particle picture and the exact numerics shrinks as we increase $\ell$ and improves at long time. In the left panel, we plot $\Delta S_A(t)$ for the same states using~\eqref{eq:dsa_jk_sum_qpp_final}. We also see the same behaviour as in the second R\'enyi asymmetry, namely a double crossing of the asymmetry curves leading to the avoidance of the quantum Mpemba effect.

\subsubsection{Power-law hopping and pairing }
\label{sec:LR_hopping_and_LR_pairing_hamilts}

\begin{figure}[t]
    \centering
   
    \includegraphics[width=.99\textwidth]{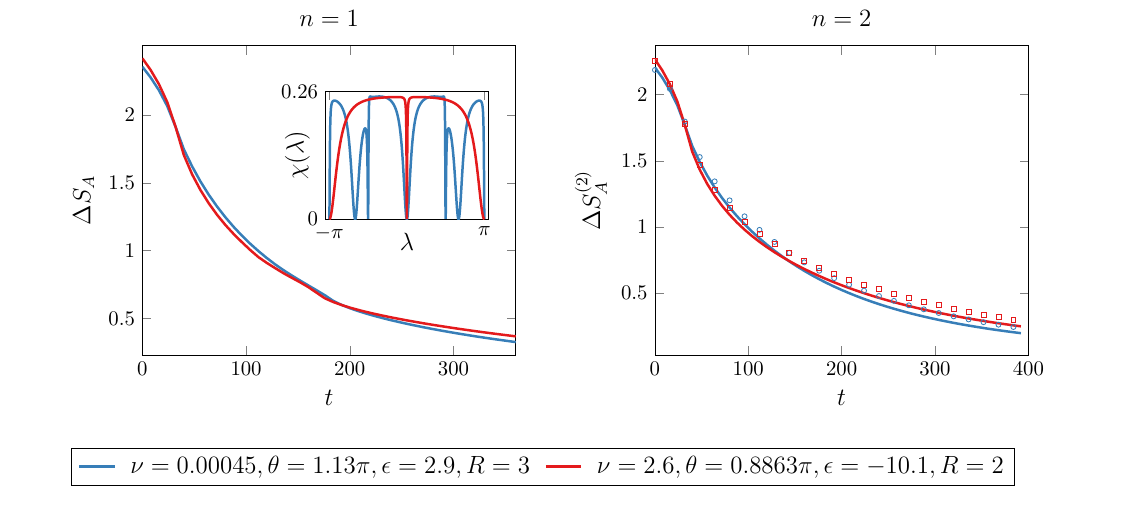} 
    \caption{ Left panel: time evolution of $\Delta S_A(t)$  obtained using Eq.~\eqref{eq:dsa_jk_sum_qpp_final} for the free-fermion chain quenched from the ground state of the long-range Hamiltonian~\eqref{eq:long_range_hopping_Ham} for two different sets of parameters with range $R=3$ (blue curves) and $R=2$ (red curves). In the inset, we represent the mode occupation of Cooper pairs in the initial states. Right panel:   evolution of $\Delta S^{(2)}_A(t)$ for the same two initial states as the left panel. The symbols are the exact value of the R\'enyi EA computed numerically using Eq.~\eqref{eq:exact_nth_Renyichargedmoment} and the continuous curves are quasi-particle prediction in Eq.~\eqref{eq:QPP_renyi_EA}. In both panels, we take as subsystem $A$ an interval of length $\ell=70$. The double crossing of the entanglement asymmetries undermines the appearence of the QME.}
    \label{fig:long_range_hopping}
\end{figure}

The second class of Hamiltonians is similar to those first considered in~\cite{vodola_njphys_2016}, having both long-range pairing and hopping with the same power-law long-range decay,
\begin{equation}\label{eq:long_range_hopping_Ham}
H_{0}=\sum_{n=1}^N\left[\cos{(\theta)} a_{n}^{\dagger}a_{n}  + \sum_{l=-R}^R\sin{(\theta)}|l|^{-\nu}\left( a_{n}^{\dagger}a_{n+l} + \frac{l}{|l|}(1+\epsilon)\left(a_n^{\dagger} a_{n+l}^{\dagger}-a_n a_{n+l}\right)\right)\right].
\end{equation}
Here, the parameter $\theta$ controls the relative strength between the chemical potential and the hopping and pairing amplitudes. We also introduce an offset $\epsilon$ which allows one to tune the relative strength of pairing and hopping.
Again we consider  a truncation in the power-law which can go up to length $R$.

In Figure~\ref{fig:long_range_hopping}, we plot both the dynamics of the entanglement asymmetry of the short-range free fermion chain~\eqref{eq:xx_hamil_real_space} initiated in the ground state of~\eqref{eq:long_range_hopping_Ham} for two different sets of parameters. One of the states (in red) corresponds to a much larger damping exponent $\nu$ than the other (in blue). As in the previous case, we plot in the right panel the dynamics of $\Delta S_A^{(2)}(t)$ using both the exact expression~\eqref{eq:exact_nth_Renyichargedmoment} and the quasi-particle prediction~\eqref{eq:QPP_renyi_EA}. In the left panel, we present the time evolution of $\Delta S_A(t)$ obtained from~\eqref{eq:dsa_jk_sum_qpp_final}. Again we find good agreement between the exact expression and the quasi-particle picture and the same qualitative behaviour in both the R\'enyi and von Neumann quantities, namely a double crossing of the $\Delta S_A(t)$ curves, which spoils the QME.

\section{Physical interpretation}
\label{sec:discussion}

\begin{figure}[t]
    \centering
    \includegraphics[width=.49\columnwidth]{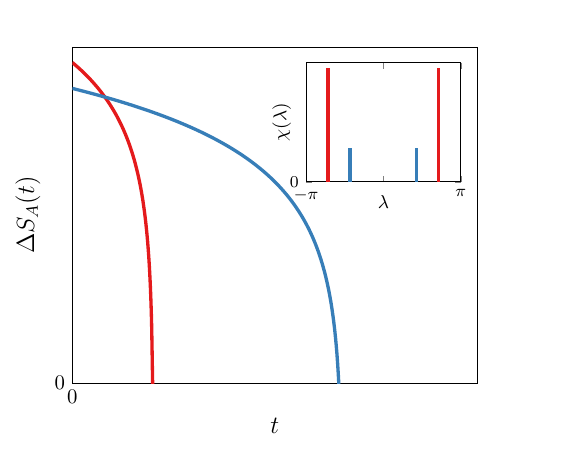} 
 \includegraphics[width=.49\columnwidth ]{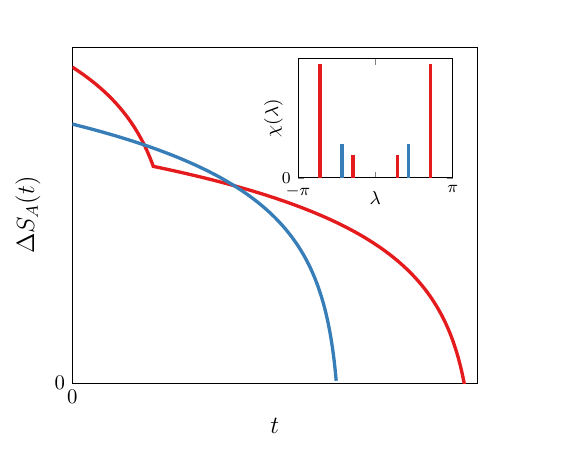}
    \caption{Sketch of the qualitative profile of $\Delta S_{A}(t)$, predicted by the saddle point approximation~\eqref{eq:spa_dsa}, after a quench to a free fermionic chain with group velocity $v(\lambda)=\lambda$ from the states with the delta comb charged susceptibilities~\eqref{eq:comb_susc} depicted in the inset of each panel.}
    \label{fig:delta_combs_toy_example}
\end{figure}

In the preceding section, we presented cases where the entanglement asymmetry exhibits a richer pattern of symmetry restoration than had previously been seen.  In this section, we investigate the underlying mechanisms for this behaviour using the quasi-particle picture.
To this end, we return to the expressions  \eqref{eq:Renyi_with_J}  and \eqref{eq:dsa_jk_sum_qpp_final} for the entanglement asymmetry written in terms of $J_k(t)$.  We concentrate on the intermediate time regime where we can evaluate $J_k(t)$ using a saddle point approximation which is valid provided  $x_\zeta(\lambda)\gg 1/\ell$ and obtain
\begin{eqnarray}
   J_{2k}(t)=\int_{-\pi}^\pi \frac{{\rm d}\alpha}{2\pi}e^{-2ik\alpha+\ell \int_{-\pi}^\pi {\rm{d}} \lambda\,  x_\zeta(\lambda) f_{\alpha}(\lambda)}\approx \frac{e^{-\frac{\left(2k-k_\zeta\right)^2}{2 \sigma_\zeta^2}}}{\sqrt{\pi \sigma_\zeta^2}}.
\end{eqnarray}
Thus, in this regime, the $J_{2k}(t)$ are approximately normally distributed with a variance and center given by
\begin{equation}
    \label{eq:sap_variance_center}
\sigma_\zeta^2=\ell\int _{-\pi}^\pi {\rm d}\lambda\,\chi(\lambda) x_\zeta(\lambda),\qquad k_\zeta=\ell\int _{-\pi}^\pi \frac{{\rm d}\lambda}{\pi}\vartheta(\lambda)x_\zeta(\lambda),
\end{equation}
where $\chi(\lambda)=\vartheta(\lambda)[1-\vartheta(\lambda)]/(2\pi)$ is the variance per quasi-particle. This variance can be interpreted as the quasi-particle charge susceptibility and a peak at certain wavevectors $\lambda$ indicates a condensation of modes at this value. Since $2\pi\chi(\lambda)=|\bra{{\rm GS}} b_{\lambda}^\dagger b_{-\lambda}^\dagger\ket{{\rm GS}}|^2$, it can also be understood as the mode occupation of Cooper pairs in the initial state, responsible for the symmetry breaking~\cite{ASYMM01murciano2023xympemba}.   Recalling that  $J_{2k+1}(t)=0$ and also from their definition that $\sum_{k\in\mathbb{Z}}J_k(t)=1$, we see that the $J_k(t)$,  in the saddle point approximation,  constitute a probability distribution.  Combining this with our earlier observation in Sec.~\ref{sec:vN_asymmetry_replica_limit} relating $J_k(t)$ to the full counting statistics, we can interpret these quantities as the probability that the complete pairs of quasi-particles contribute the value $k$ to the charge measured (from half filling) in the subsystem.  With this interpretation, it is now evident that $J_{2k+1}(t)$ should vanish since pairs can contribute only even charge.  
Plugging this into~\eqref{eq:dsa_jk_sum_qpp_final}, we find that 
\begin{equation}
    \label{eq:spa_dsa}
    \Delta S_{A}(t)\approx\frac{1}{2}\log{(\pi\sigma^{2}_{\zeta})}+\frac{1}{2}.
\end{equation}
Hence, in this regime,  $\Delta S_{A}(t)$ is the Shannon entropy of the probability distribution of measuring the number of quasi-particles that belong in complete pairs within $A$ and is completely determined by the variance $\sigma_\zeta^2$.   The existence or not of the QME can therefore be inferred, at least within the validity of the saddle point approximation, by examining the behaviour of $\sigma^{2}_{\zeta}$.  We note that the interpretation of  $J_k(t)$ as a probability does not extend beyond the regime of validity of the saddle point approximation as, in the long time limit $t\gg \ell$, $J_k(t)$ can be negative~\cite{ASYMM08rylands2023microscopic}.

To understand the behaviour of $\sigma_\zeta^2$, it is instructive to examine a toy example.  Let us consider a quench from two initial states whose occupation functions are strongly peaked about a finite number of modes.   We take them to be such that the quasi-particle variance has the structure of a delta comb,
\begin{equation}\label{eq:comb_susc}
\chi_{1,2}(\lambda)=c^{1,2}_{+}\delta(|\lambda|-\lambda^{1,2}_+)+c^{1,2}_{-}\delta(|\lambda|-\lambda^{1,2}_-),
\end{equation}
where $\lambda_{\pm}^{1,2}$ are the occupied modes  of each states and $c_{\pm}^{1,2}\in[0,1/4]$ are constants specifying the variance of each mode. The different states are indexed by the superscript $1,2$ and the modes by  the subscript $\pm$.  These states are depicted in the left panel of Figure~\ref{fig:delta_combs_toy_example}.   For simplicity, we also take the quasi-particle velocity to be linear, $v(\lambda)=\lambda$. The variance of each state is then straightforward to evaluate
\begin{eqnarray}
\sigma^2_\zeta=2c_{+}^{1,2}x_{\zeta}(\lambda^{1,2}_{+})+2c_{-}^{1,2}x_{\zeta}(\lambda^{1,2}_{-}).
\end{eqnarray}
When only a single mode is occupied, i.e. $c_-^{1,2}=0$, we have that $\sigma_{\zeta}^2$ decays linearly in time with slope $-2|v(\lambda^{1,2}_{+})|$.  By appropriately tuning the constants, $c^1_+>c^2_+,~\lambda_+^1>\lambda_+^2$, it is possible to engineer a single crossing of the variance and hence $\Delta S_A(t)$, as shown in the left panel of Figure~\ref{fig:delta_combs_toy_example}.  If instead we allow $c^{1,2}_-$ to be nonzero as well, then the variance again decays linearly but has a change of slope at time $t=\ell/(2 |v(\lambda_+^{1,2})|)$.  At this point, the modes $\lambda_{+}^{1,2}$ have had enough time to traverse half of the subsystem meaning that one member has exited $A$, the pair becomes incomplete and so only  the modes $\lambda_-^{1,2}$ contribute to the entanglement asymmetry.  This multi-slope behaviour of the variance causes the entanglement asymmetry to have a richer structure and allows for the two states to exhibit multiple crossings throughout their evolution, as in the right panel of Figure~\ref{fig:delta_combs_toy_example}.  

From this simple example, we see that the QME can occur if the two states have occupation functions such that the density of Cooper pairs, $\vartheta(\lambda)(1-\vartheta(\lambda))/2\pi$, is  strongly peaked about only a single quasi-particle pair.        If, on the other hand, the occupation functions have a more complicated structure, which translates to $\chi(\lambda)$, then the entanglement asymmetry can have  richer dynamics including multiple crossings that could lead to the QME or the avoidance of it. 

The delta combs are the most extreme example of wavevector population modulation since we have only certain excited modes in the system, but one can imagine less extreme situations, with smoother multiple peaks.
In particular, we can observe in the inset of the left panel of Figure~\ref{fig:xy} that for the single crossing of the XY chain ground states, the quasi-particle variance $\chi_{1,2}(\lambda)$ have only two symmetric peaks.  We can contrast this with the same quantity for the ground states of Hamiltonians with long-range terms depicted in the left panel insets of Figures~\ref{fig:dsa_qpp_double_crossing} and~\ref{fig:long_range_hopping}.  Here we see that, in the state corresponding  to a smaller $\alpha$ and hence longer range pairing, $\chi(\lambda)$ has a much more complicated profile. As was anticipated in Section~\ref{sec:corrmatr_of_the_ground_states},
the long-range terms in $H_{0}$ yield an oscillating density of occupied modes $\vartheta(\lambda)$ that translates into a density of Cooper pairs $\chi(\lambda)$ with multiple peaks.
This concentration of Cooper pairs about certain wavevectors endows $\Delta S_A(t)$ with a multi-slope structure thereby allowing for the avoidance of the QME despite an initial crossing.

\section{Conclusions}
\label{sec:Conclusions}
In this paper, we have studied the dynamical restoration of particle number symmetry in the short-range free fermion chain quenched from ground states of long-range Kitaev models. To do this, we have employed the entanglement asymmetry which measures the amount of symmetry breaking at the level of a subsystem using entanglement entropy based metrics. This quantity has previously been used to identify a curious phenomenon called the quantum Mpemba effect in which non-equilibrium states of a closed quantum system can relax faster if they are initially further from equilibrium. This effect is manifest in the entanglement asymmetry as inverting of the hierarchy of the asymmetry for two different states between short and long times, i.e. if state 1 initially has higher entanglement asymmetry than state 2 then the quantum Mpemba effect describes the situation where, at long times, state 2 has higher asymmetry than state 1. Up till now, in all cases studied the effect occurs with just a single crossing of the asymmetry curves at times $\tau_M\sim \ell$. In this work, we have shown that this scenario is not the only possibility and the dynamics of the entanglement asymmetry can have much richer phenomenology. 

By preparing the system in the ground state of a long range free fermion Hamiltonian, we have shown that the asymmetry can exhibit an avoidance of the quantum Mpemba effect characterized by an initial inverting of the relative asymmetry for two such states followed at a later time by a reversion to the initial ordering. This phenomenology can be understood using an approximate, intermediate time description which relates the entanglement asymmetry to the Cooper pair density or charge susceptibility. In this picture, the quantum Mpemba effect can be seen as arising from the concentration of Cooper pairs at different momenta between these two states. For states which are ground states of long range models, the more complicated nature of the models endows the occupation functions with a much more intricate structure, which in turn translates into the presence of multiple peaks in the Cooper pair density in momentum space. Ultimately, this leads to a richer dynamical interplay between the two states. While we have investigated only certain areas of the parameter space of these models, we have found that avoidance quantum Mpemba effects are not necessarily rare events. Furthermore, we expect that a higher number of crossings can be found following a more thorough search of the parameter space.

Our work highlights two important points concerning the dynamics of the entanglement asymmetry and the quantum Mpemba effect. First, the existence of the quantum Mpemba effect cannot be inferred from a short time expansion about the initial state. As we have seen, while it is possible that a crossing may appear within the short time regime, subsequent crossings may appear at a later time leading to normal relaxation i.e. the more symmetric state relaxing faster. The QME necessarily comes from comparing the $t/\ell=0$ and $t/\ell\to \infty$ behaviour of the system. Second, if one restricts to a study of the initial and final times, the rich intermediate time dynamics involving the interplay of relative slower and faster relaxation may be missed.  

Although we have studied only quenches to a non-interacting system, we expect that similar behaviour can be seen also in interacting integrable models, which also admit a quasi-particle description. Such models can have many species of quasi-particles with different velocities and so similar behaviour could possibly be witnessed in relatively simpler initial states. For example, the Hubbard model and Gaudin-Yang models, quenched from certain initial states exhibit a clear two-slope structure in the entanglement entropy reminiscent of the toy example discussed above~\cite{rylands2022integrable,rylands2023solution}.

\textbf{Acknowledgements}
This work has been supported by the European Research Council under Consolidator Grant number 771536 ``NEMO''. We thank J. De Nardis, S. Fraenkel, D. X. Horvath, S. Murciano, G. Policastro and  S. Yamashika for useful discussions.

\newpage
%\section{References}
%\bibliographystyle{unsrt}

\end{document}